# ARTICLE

# Enhancement in electromechanical properties of piezoelectric thin film through strain-induced domain alignment

Antony Jeyaseelan A,[a] Sandip Bysakh,[b] Jalaja M A[a] and Soma Dutta *[a]

Abstract: This paper reports the impact of process-dependent structural deformation and lattice strain by doping, resulting in domain re-orientation along the a-axis. For this investigation, the smaller $La^{3+}$ cation is introduced at A-site and the longitudinal and transverse piezocoefficient properties have been studied in $Pb(Zr,Ti)O_3$ (PZT) film. Introducing smaller cations at the A-site leads to a reduction in the lattice parameter and improves the lattice matching with the Pt substrate. The XRD and HRTEM studies evidence this occurrence in both films. The HRTEM analysis also reveals the 30° long-range ordered domain alignment due to the lattice mismatch and 0° match domain alignment with the substrate in PZT and PLZT films respectively. The strain-induced 30° domain alignment in PZT enhances the longitudinal ($d_{33d}$), whereas 0° domain alignment in PLZT enhances the transverse ($d_{31}$) piezocoefficient properties. Incorporating 8% of La in the PZT lattice leads to a two-fold increase in the $d_{31}$ value compared to PZT film.

## Introduction

For the investigation of crystal structure and domain orientation in thin film structure, it is essential to identify the underlying intrinsic physical properties such as substrate lattice orientation, residual strain etc[1]. The lattice strain-induced domain/domain wall engineering can effectively tune various properties (like electronic, optical, and mechanical) of nanocrystalline thin films of a few nanometers thicknesses[2-4]. Lattice strain of thin films on their surfaces or interfaces is considered to be one of the important fundamental parameters for inducing novel properties as compared to their bulk counterparts[5-8]. The positive as well as the negative strain can be induced in the crystal lattice by changing its atomic arrangements through fine-tuning of the process parameters or by external thermal activation. The domain structure and stability of the domain are controlled by the defects, bonding and screening conditions at the interface, which can significantly affect the domain structure and in turn, the stability of specific domains[9-13]. Thus, it is possible to design or engineer a domain structure specific to particular properties or applications. In epitaxial films, the epitaxial strain arises from the matching of the in-plane lattice parameters of both the substrate and ferroelectric film[14-16] and in non-epitaxial films, the strain develops from the difference in the lattice parameters of the film and substrate, which may be an order of magnitude higher than that of the epitaxial strain. The crystal lattice's atomic arrangement, i.e., long-range order (single domain) or short-range order (multi-domain), can be customised by changing the process conditions or adding dopants[17-19]. The electromechanical properties of piezoelectric films strongly depend on the domain structure/arrangements and the inbuilt strain[20]. Hence, it is worth pursuing the behaviour of piezoelectric thin films in the context of the strain effect.

Among piezoelectric materials, Lead Zirconate Titanate shows relatively large mechanical strain which can give the longitudinal piezoelectric coefficient ($d_{33}$) values as high as 700 pm/V reported elsewhere [17,20-24]. The transverse piezoelectric coefficient $d_{31}$ values reported for Lead Zirconate Titanate thick film (thickness ~ 1μm) are found to lie in the range of –28 to –70 pC/ N [24-27]. Literature reporting $d_{31}$ values of pure and La-doped PZT thin films are scant [20]. The strain field effects are more prominent in thin film structures where substrate clamping and domain orientation play a vital role towards electromechanical properties. The present study focuses on the investigation of piezo coefficients of $PbZr_{0.52}Ti_{0.48}O_3$ (PZT) and $Pb_{0.92}La_{0.08}(Zr_{0.52}Ti_{0.48})_{0.98}O_3$ (PLZT) films to understand the correlation between lattice misfit strain on domain orientation and its electromechanical behaviour.

## Results and discussion

[a] Materials Science Division, Council of Scientific and Industrial Research, National Aerospace Laboratories, Bangalore – 560017, India
[b] Transmission Electron Microscopy Laboratory, Council of Scientific and Industrial Research, Central Glass and Ceramic Research Institute, Kolkata – 700032, India

To examine the changes in lattice parameters and misfit strain due to doping, x-ray diffraction (XRD) studies were carried out on pure and La-doped PZT films. Figure 1 shows the XRD pattern of PZT and PLZT films deposited on Pt(111)/SiO$_2$/Si. Both the films exhibit peaks along (100), (110), (111), (200), (201), and (211) which correspond to P4mm perovskite phases. Figure 1 displays the XRD pattern of pure and La adoped PZT films deposited on Pt(111)/SiO$_2$/Si. Both the films exhibit high-intensity peaks along (100), (110), (111), (200), (201), and (211) which correspond to perovskite phases. However, a comparative study with the diffraction peaks of bulk PZT and PLZT (JCPDS card no: 33-0784 and 29-0776, respectively) shows a delineated shift in the peak positions in both the thin films. Such a shift in peak positions suggests that the films are strained.

**Table 1:** Refined lattice parameter obtained after Rietveld refinement of XRD diffraction peaks

| Samples | Lattice parameters | | $R_p$ | $R_{wp}$ | $R_{Exp}$ | $\chi^2$ | $R_f$ |
|---|---|---|---|---|---|---|---|
| PZT | a=b=4.0521, c=4.1080 | c/a =1.0138, V=67.45 | 28.3 | 30.3 | 24.7 | 1.50 | 8.31 |
| PLZT | a=b=4.0360 c=4.0754 | c/a =1.0097, V=66.36 | 30 | 30.9 | 24.4 | 1.60 | 8.24 |
| (PZT) 033-0784 | a=b=4.0360, c=4.1460 | c/a=1.0273, V=67.535 | | | | | |

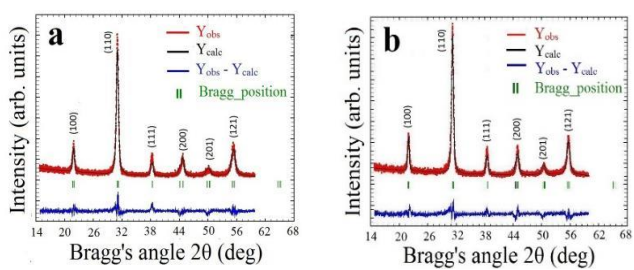

**Figure 1.** Observed (red line), calculated (black line), Bragg position (green line) and difference (blue line) profiles obtained after Rietveld refinement of XRD data of (a) PZT, (b) PLZT films

Rietveld refinement plots for both the PZT and PLZT films are shown in Fig. 1a and 1b. A close match between the theoretically calculated and the observed diffraction patterns of the tetragonal PZT(t-PZT), belonging to the P4mm space group is observed. The refinement parameters, i.e., profile ($R_p$), weighted profile ($R_{WP}$), Bragg(R), and expected R factors ($R_{exp}$) are given in Table 1. The calculated values of ($R_{wp}/R_{exp}$) for PZT and PLZT are 1.226 and 1.251 respectively, which indicates the correctness of the fits. The marginal difference in the structural parameters between PZT and PLZT films may be due to small variations in the position of the atoms. Additionally, the experimental ($Y_{obs}$) and theoretical ($Y_{calc}$) profiles display minor differences in the intensity scale near zero, as shown by the ($Y_{Obs}$–$Y_{Calc}$) line in Fig. 1.

A close look at the (100) and (110) peaks shows a peak shift in $2\theta_{100}$ (22.16 to 22.20) and $2\theta_{110}$ (31.28 to 31.45) degrees was observed in PZT and PLZT films. The lattice volume contraction in the PLZT could be due to the substitution of smaller La$^{3+}$ with larger Pb$^{2+}$ ions. The a-axis texturing of the PLZT is quantitatively calculated from the integrated intensities of the fitted curve using equation 1 and found to be 33% which is two times higher than that in PZT (17.8%).

$$P_{(100)} = \frac{I(100) \times 100}{I(100)+I(110)} \quad (1)$$

In PZT film, the in-plane compressive stress arises because of the greater difference in the lattice parameters with cubic Platinum (Pt) (a=b=c=3.92) and thereby shows a higher tendency of preferential crystal growth along the c-axis. Due to the compositional change with La doping in PLZT, the cell volume and **c/a** ratio decreases. The decrease in the **c/a** ratio implies that the film experiences less compressive stress compared to that in the PZT, which promotes the growth of the film along the **a-b** plane. Hence, a structural distortion takes place within the film because of the constraint imposed by the substrate as well as lattice mismatch. It is believed that such conditions modify the TiO$_6$ octahedral environment with different crystal geometries. The combined effect of misfit strain and grain orientation (strain developed due to the difference in lattice parameters between the film and the substrate) changes the cell dimension and in turn, the domain orientation in PZT film along c-axis and along a-axis for PLZT. It is known that lattice expansion in a plane should accompany shrinkage of the lattice dimension perpendicular to the plane. Hence, the lattice expansion cannot be explained in terms of simple elastic misfit strain alone. The misfit strain can be calculated using the following equation 2,

$$\varepsilon_m = \frac{(a_s - a_f)}{a_s} \quad (2)$$

where '$a_s$' is the lattice parameter of the substrate and '$a_f$' the lattice parameter (along x-axis) of the film.

When the lattice parameters of PZT (4.052) and PLZT (4.035) are compared with the lattice parameter of Pt(3.92) substrate, the misfit strain value is found to be around -0.034 for PZT and -0.029 for PLZT. The negative sign implies compressive strain. The magnitude of the misfit strain suggests that compressive strain is reduced in PLZT, and that explains why the c/a (tetragonality factor ~1) reduces in PLZT.

### 1.1. HR-TEM analysis

Transmission Electron Microscopy (TEM) and Energy Dispersive X-ray (EDX) spectroscopy studies were carried out

to correlate the properties of the films with their fine-scale microstructure and chemistry.

The bright-field TEM (BF-TEM) image in Fig.2a shows well-grown grains in the PZT film, which is having good adherence on the Pt electrode. However, the presence of both the lattice strain and the scattered nano-porosity contribute to the lattice defects in the PZT crystals. The perpendicularity of grain boundaries with respect to the surface of the Pt substrate suggests a tendency toward the preferred direction of grain growth. The selected area diffraction pattern (SAED) shown in Fig.2b was taken by selecting one particular PZT crystal along with the underlying Pt layer. The pattern shows [110] crystallographic zone axis of the Pt grain and the [100] direction PZT along incident electron beam direction during HRTEM imaging condition.

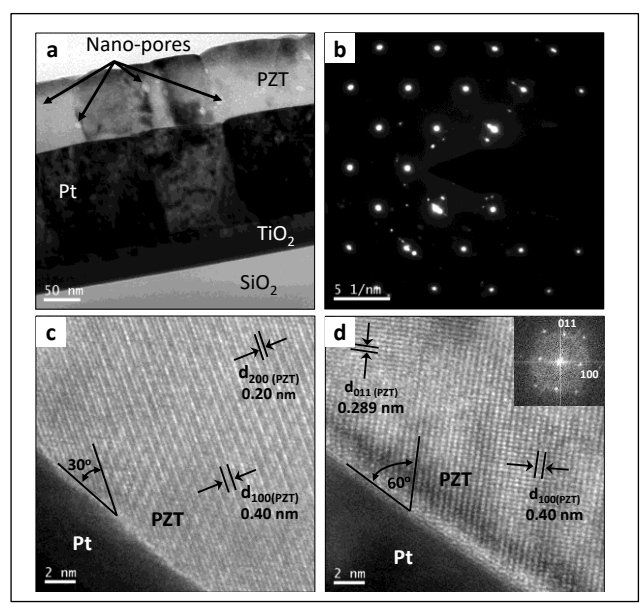

**Figure 2.** (a) HRTEM bright field (b) SAED patterns of PZT thin film, (c) HRTEM image of Pt-PZT interface showing 30° angle between substrate and (100) lattice planes in a PZT grain, and (d) (011) and (100) lattice fringes at 60° angle with the substrate in another PZT grain and Pt crystal.

High-Resolution Transmission Electron Microscopy (HRTEM) imaging was performed to obtain finer details at the PZT/Pt interface. The HRTEM images in Fig.2 show a highly adherent interface between the Pt and PZT grains with no evidence of any reaction product at the interface. As can be seen from the HRTEM images (Fig 2c and 2d), the PZT crystal orientations are different in the two images. Such varied grain orientation was observed all over the PZT film. This appears to be due to the influence of the varied orientation of the Pt crystals in the underneath layer The HRTEM image in Fig.2c the PZT lattice fringes with alternately varying intensity can be observed. These lattice fringe spacings were measured and are realized to be the 100 (0.4 nm) and the 200 (0.2 nm) superlattice 'd-spacing' of the t-PZT as indicated in Fig. 2c. These spacing values correspond to a t-PZT phase [JCPDS card no. 14-0031]. The angle between these lattice fringes and the underlying Pt-layer surface was measured to be about 30° as shown in Fig.2(c). This means that the 'a-axis' direction of the PZT crystal is oriented at an angle of 30° away from the film surface normal. This is the best-observed orientation of 100 planes in the t-PZT film.

The HRTEM image in Fig.2d shows the lattice fringes in another PZT grain showing grain orientations different from that in Fig.2c. Measurement of the lattice fringe spacings in the HRTEM image in Fig.2d suggests that the two mutually perpendicular sets of lattice fringes correspond to (100) and (011) crystallographic planes of t-PZT [JCPDS card no. 33-0784]. This HRTEM image, as in Fig.2d, represents another one of the varied crystal orientations of the t-PZT crystals. In this image, the electron beam is parallel to the $(0\bar{1}1)$ crystallographic zone axis of t-PZT, and to the [110] direction of the Pt crystal during HRTEM imaging condition. It can be observed that the fringes of the (100) lattice planes are making an angle of about 60° with the substrate plane. This means that the 'a-axis' direction of this PZT crystal is lying at an angle of 60°, which is quite away from the film surface normal. The Fast Fourier Transform of this HRTEM image in the inset shows the zone axis pattern which has been indexed showing (100) and (011) planes of the t-PZT crystal.

This TEM study of the PZT film shows that there is no definite orientation relationship between the Pt and PZT crystals. Instead, the t-PZT grains show varied orientation relations, if any, with the Pt grains underneath. Hence, an inference can be drawn that the crystallographic orientation of the PZT grains is much less influenced by the grain orientation of the Pt substrate than that by the misfit strain. In the case of this PZT film, the best orientation of the 'a-axis' was found to be at an angle of 30° to the film surface normal.

The HRTEM studies on the cross-section of the PLZT thin film too revealed similar results. The BF-TEM image of the PLZT film-substrate cross-section in Fig.3a showed growth of the PLZT film with a fairly uniform thickness of about 100nm over the Pt substrate. The SAED pattern in Fig. 3b shows the strong diffraction spots from the Pt substrate and the weaker spots from a PLZT grain.

The HRTEM images presented in Fig.3c and 3d show the PLZT/Pt interface at two different locations. At these locations, the orientations of the PLZT crystals are different in respect of the underlying Pt crystals. This suggests that there is little influence of the substrate Pt grain orientation on the PLZT grains. The HRTEM image in Fig.3c shows the (100) and the (200) superlattice fringes of PLZT crystal oriented exactly parallel to the substrate surface. This means that in this PLZT grain, the 'a-axis' orientation is perpendicular to the substrate surface and is the desired orientation.

Figure 3d is the HRTEM image of PLZT grains grown on a Pt electrode. The image shows that the two adjacent PLZT grains containing (110) lattice fringes of 0.29 nm spacing have a mutual orientation relation similar to a crystallographic twin relationship. It may be noted that these PLZT grains have assumed the tetragonal lattice parameters [PDF no. 29-0776], in agreement with our XRD results. This HRTEM image in Fig.3d represents one of the other varied crystal orientations in the PLZT film.

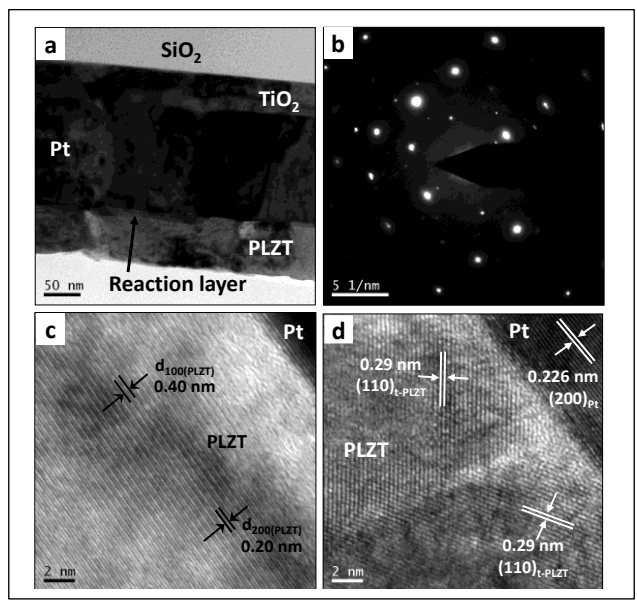

covalent B–O bonds, the internal vibrational Raman shifts are observed at higher frequency regions whereas those corresponding to A-O bonds appear at lower frequency regions[33]. The effect of substitution of ($La^{3+}$) cation at A-sites was analyzed from the room temperature Raman spectra of PZT and PLZT films (Fig. 4).

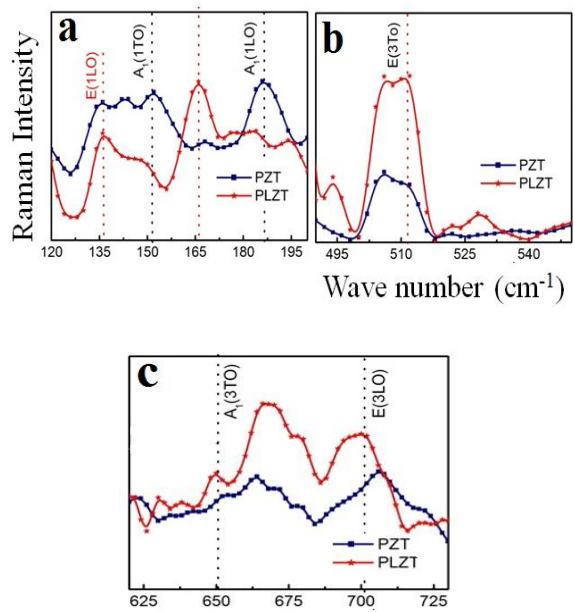

**Figure 3.** Bright-field TEM images of (a) PLZT 100nm film on platinised silicon wafer, (b) SAD pattern of PLZT film (c) lattice fringes in PLZT grain showing 100 and 200 PLZT lattice planes on Pt crystal, and (d) Pt- PLZT interface showing 110 lattice fringes in differently oriented PLZT crystals

Based on the HRTEM results discussed above, it may be concluded that in both the PZT and the PLZT films, the grains grown on the Pt electrode have preferably assumed the respective tetragonal crystal structures. The grains in PLZT film showed a more favourable 'a-axis' orientation, i.e., perpendicular to the substrate, whereas, the best-observed orientation in PZT film has been found to be at an angle of 30° away from the ideal orientation perpendicular to the substrate. Thus, better electro-mechanical properties were exhibited by the PLZT films as will be presented in the following sub-sections.

Raman scattering is a powerful technique to understand the coordination of local symmetry, and it provides insights into the molecular vibrations and distortions in the crystal lattice[30,31]. The Raman backscattering method was used for the local structure analysis of the PZT and PLZT thin films. In general, the $ABO_3$ perovskite system is grouped into internal and external modes of vibration in Raman studies. In an $ABO_3$ perovskite system, the B-site cations form covalent bonds with oxygen atoms. The A-O lattice vibration is grouped under the external mode of lattice vibration whereas B-O is grouped under the internal mode of vibration. With the stronger

**Figure 4.** Raman spectrum for PZT and PLZT films (a) low-frequency region, (b) mid-frequency region, and (c) high-frequency region; Red and blue lines indicate PZT and PLZT respectively

Figure 4a represents the three Raman active modes E(1LO), $A_1$(1TO), and $A_1$(1LO) at 135, 152 and 165cm$^{-1}$ respectively, corresponding to the symmetric stretching of Pb-O and La-O bond of PZT and PLZT films. The intense band at E(1LO) and $A_1$(1TO) reveals that the Pb-O stretching is symmetric in PZT film and asymmetric in PLZT. The asymmetry in Pb-O stretching in PLZT arises because of the interference between the La-O and the Pb-O bonds. The Raman spectra shows high intense shoulder-type peak around 510cm$^{-1}$ region in PLZT corresponds to in-plane mode of vibrations of Ti-O bond[34]. The two Raman modes at $A_1$(3TO) (649 cm$^{-1}$), and E(3LO)(699 cm$^{-1}$) also support this phenomena. Thus this Raman studies support that the in-plane stress is relieved by incorporating $La^{3+}$ ion doping in PZT film, which is well collaborate with the misfit strain value by XRD and domain re-orientation by TEM analysis.

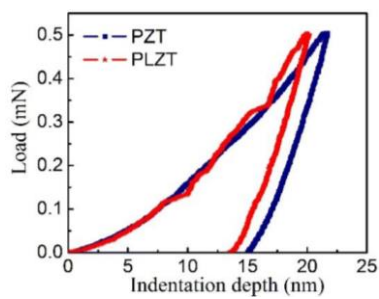

**Fig.5.** Plot of nanoindentation depth vs load for PZT and PLZT films

Mechanical deformation in PZT was reported earlier in the context of grain boundary and porosity of nanocrystalline material[32]. In this report, an attempt has been made to establish the relationship of mechanical deformation with lattice strain, if any. The mechanical deformation in the PZT and PLZT films was examined using a loading-unloading nano-indentation curve. No discontinuity in elastic deformation was observed in the loading-unloading curve of PZT film, taken at different locations throughout its thickness, whereas, discontinuity was observed in the case of PLZT film. It is believed that the non-uniform distribution of micro-strain is the reason for the observed discontinuity in the loading-unloading curve. However, it suggests the grain growth in different directions which could increase the hardness. The hardness values of PZT and PLZT thin films at 0.5mN load were measured to be 9.86 and 10.35GPa respectively and its comparable to those reported [35, 36,37]. The Young's modulus of PZT and PLZT thin films calculated from the nano-indentation curve (Fig.5) is in the order of 135GPa and 149GPa respectively. The higher magnitude of compressive strain in PZT lattice sets the domains to align along the c-axis with minimum or no domain walls. In the favoured domain orientation, the whole crystal lattice behaves like an anisotropic single crystal with long-range ordering that reduces the mechanical strength of PZT as reflected by the hardness value. The lattice strain plays an important role in thin film structure to control mechanical behaviour.

**1.2. Electromechanical and ferroelectric properties**

The small signal piezo-phase angle measurements vs applied voltage were carried out to explore the extrinsic contribution. The domain alignment of a-axis or c-axis orientation of thin films is strongly dependent on the grain boundary and lattice strain state(Fig. 6a).

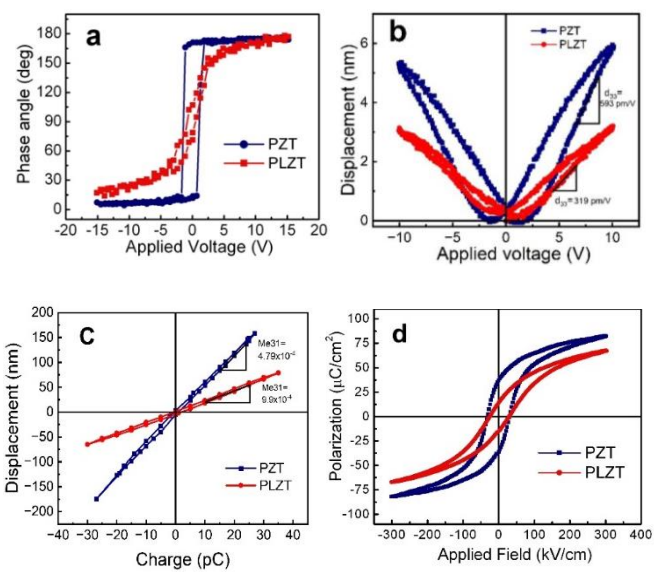

**Figure 6.** (a) Represents the 180° and non-180° domain orientation of PZT and PLZT with respect to the applied voltage (b) piezo butterfly loops (c) charge vs displacement plot and (d) field dependent remnant polarization for both PZT and PLZT films

The phase angle switching plot with respect to applied voltage explains why local 180° domain switching is favoured in PZT and non-180° switching occurred along with 180° domains in PLZT film. The hindrance of 180° domain switching in PLZT film is due to the relaxed compressive strain and the presence of short-range order where grain boundaries prevent the local domain movement. The ideal z-type anisotropic loop of PZT identifies the complete domain switching of 180°, which in turn suggests a significant amount of long-range ordered grains in PZT film. In the case of a-axis oriented isotropic PLZT film, the linear behaviour s-type loop suggests the gradual multi-angle short range order domain switching of the differently oriented domains (Fig. 6a).

The large signal piezoelectric coefficients $d_{33}$ and $d_{31}$ were measured to establish the correlation between the mechanical, and electromechanical properties of both the films (Table 2). The piezoelectric butterfly loops for both PZT and PLZT are shown in Fig.6b. The slope of the voltage-dependent displacement gives the effective $d_{33}$ value. The PZT sample shows a higher effective $d_{33}$ value (593 pm/V) compared to PLZT (319 pm/V). The $d_{33}$ values of PZT thin film are comparable with the values reported in PZT (with composition 52/48) thick films (~1 micron) [24,25]. The $d_{33}$ value in PZT thin film is found to be higher than the solution-processed PZT thick films reported previously and comparable [26]. The $d_{33}$ value in PLZT is three-fold higher than Ba-Substituted PZT [27]. Figure 6(c) shows the charge vs displacement plot for both PZT and PLZT films. $Me_{31}$ values are obtained from the slope of the charge vs displacement plot and are directly proportional to the $d_{31}$ of the measured films. The $d_{31}$ is then calculated using equation 3.

$$d_{31} = Me_{31} \cdot \frac{l_1^2}{4Ah(1-v_s)} \quad \text{(3)}$$

where $l$1 is the distance between the lower $A$, is the area of the top electrical contact, h is the thickness of the sample and $v$ is the Poisson ratio of the substrate. The $Me_{31}$ and $d_{31}$ values of both PZT and PLZT films are tabulated in Table 2 and $d_{31}$ value of PLZT film is found to be (77pC/N), which is two-fold higher than that of PZT (33pC/N) film. The two-fold increase in d$_{31}$ value in PLZT is due to the preferred 'a' axis orientation of the film. From the experimental results, it is understood that the incorporation of smaller La$^{3+}$ ions in A-site leads to a decrease in misfit strain and improves the transverse piezocoefficient properties of PZT film. The domain alignment depends on the crystal structure and the geometry of TiO$_6$ octahedral environment of the film which is mainly decided by the substrate-film clamping state and thereby generated stresses. When the bias field aligns, all the dielectric domains parallel to its direction (180º domain movement), and the stretching of TiO$_6$ octahedral bond gives the ultimate displacement in the unit cell which is observed in PZT. The TiO$_6$ octahedral environment plays the major role in producing 180º domains in PZT and non-180º domains in PLZT (Fig.7c). It is understood that the misfit strain decides the growth and orientation of thin film as observed in the present study by the c-axis orientation in PZT and a-axis orientation in PLZT film.

The higher d$_{33}$ (Fig. 6b) in the c-axis oriented PZT film suggests that the compressively strained film is easy to polarize along the c-axis (out of plane). The cumulative effects of unidirectional dielectric domains contribute towards higher displacement. On the other hand, when the grains in PLZT film get aligned along the a-axis due to reduced compressive strain, the d$_{33}$ value becomes low due to the hindrance in dipolar orientation. That is why when the hardness of the film is high; it is constrained to give low displacement. As the higher hardness value was observed for PLZT in the nanoindentation test, ideally a low displacement was probable for PLZT. The movement of the B-site atom in ABO$_3$ (TiO$_6$) is the basis of the expansion of the lattice when the electric field is applied to a piezoelectric material. The B-site TiO$_6$ environment contains two types of Ti-O bond, i.e., axial Ti-O bond along the c-axis and equatorial Ti-O bond along the a-direction. The piezoelectric properties depend on the flexibilities of these bonds. In PZT film, due to higher compressive strain, the Ti-O bond in the equatorial position shortens and axial Ti-O bond elongates along the c-axis. It exaggerates the tetragonal distortion and aligns the domains to switch parallel to the applied field and helps in attaining higher displacement in PZT. In the case of PLZT the relaxation in the compressive stress along in-plane direction leads to the formation of random non-180° domains which can contribute towards the piezoelectric d$_{31}$ coefficient, it is expected to observe a higher d$_{31}$ value in PLZT.

Figure 6(d) shows the ferroelectric hysteresis loop of PZT and PLZT films. The main observations from the plot are (i) the occurrence of domain switching and ferroelectric field reversal, (ii) a slimmer pattern of the PLZT loop than PZT, and (iii) higher remnant polarization of PZT film than PLZT. As highlighted in the previous section, the higher remnant polarization in PZT (37.30µC/cm$^2$) is due to the more aligned 180º ferroelectric domains along the c-axis resulting from the higher compressive strain. The gradual linear increase in the polarization in PLZT film with the applied field is due to the slow polarization switching of unfavored domains.

| Film | Mechanical properties (GPa) | | Misfit strain | Domain orientation | Piezoelectric properties | | | |
|---|---|---|---|---|---|---|---|---|
| | | | | | Longitudinal mode | Transverse mode | | |
| | Hardness | Young's modulus | | | d$_{33}$(pm/V) | Me$_{31}$(C/m) | e$_{31}$(C/m$^2$) | d$_{31}$(pC/N) |
| PZT | 9.86 | 135 | 0.0092 | 30 to 60° | 593 | 4.79 x 10$^{-4}$ | -04.62 | -33 |
| PLZT | 10.35 | 149 | 0.0053 | 0° | 319 | 9.9 x 10$^{-4}$ | -11.45 | -77 |

## Experimental

PZT and PLZT thin films were deposited on a platinized silicon wafer (Pt (111)/Ti/SiO$_2$/Si) by chemical solution deposition method[26,28,29] followed by pyrolysis at 400°C. The films were then annealed by a rapid thermal process at 650°C. Multiple iterations of spin coating and heat process were carried out to build a thickness of 400 nm. The Metal-Insulator-Metal structure was fabricated with Au and Pt as top and bottom contact respectively.

The phase formation, surface texture and morphology of the thin films were checked by X-ray diffraction (PANalytical X'pert PRO machine, CuKα (λ = 1.604Å), and high-resolution transmission electron microscopy (TEM; Tecnai F30, FEI, Hillsboro, OR, USA).To prepare the cross-section specimens from the thin films for TEM studies, rectangular pieces of film-coated substrates were sliced by using a low-speed precision diamond saw (Buheler Inc., USA) to make 0.25mm thick discs containing the cross-sectioned sandwich joint along its diameter and fixed with G1 epoxy (Gatan Inc, USA). Then grinding and dimpling were done to bring down the specimen thickness to a residual value of 100μm. Finally, the specimen was transferred to GATAN Precision Ion Polishing System (PIPS) (GATAN Inc., USA) for argon ion polishing at 4 keV and 4° ion-beam incidence angle on both sides with double-beam modulation till perforation of electron transparency is obtained.

To study the mechanical properties of the films, the nanoindentation technique was employed. The ferroelectric and piezoelectric characteristics (at room temperature) of the PZT and PLZT films were investigated using a thin film analyzer (TF 2000E, aixACCT, Germany). The $d_{31}$ values of the films were determined through a 4-point bending setup attached to aixACCT system, where the displacement was measured under a constant mechanical load with varying electric fields. The generated charge and the displacement of the film with applied voltage were measured by a laser interferometer. From the slope ($Me_{31}$) of the linear charge displacement plot, the $d_{31}$ was calculated for PZT and PLZT thin films. To measure the $d_{33}$ the thin film analyser was used in a double-beam laser mode.

## Conclusions

In summary, it is evident that the substitution of smaller $La^{3+}$ cation in the A-site shows relaxation behaviour in the in-plane misfit strain of PZT film. The XRD analysis confirms the achievement of lower misfit strain with Pt substrate in PLZT compared to PZT film. The HRTEM analysis evident the misfit strain on the domain alignment of PZT and PLZT and it was found to be 30° and 0° misaligned with normal to the substrate. The Raman analysis also confirms the increase in the transverse vibrational mode of the $TiO_6$ octahedral environment in PLZT. The higher remanent polarization and sharp switching piezo-phase angle confirm the contribution of the c-domain (180°) in PZT film. The highly compressed in-plane stress on PZT film shows high $d_{33}$ whereas significant increase in $d_{31}$ value has been achieved of PLZT film. This anomaly in the transverse piezocoefficient value is due to the relaxation of in-plane stress with La doping in PLZT film.

## Author Contributions

A.A.J prepared the thin films, performed Ferroelectric and Piezoelectric experiments of thin films, handled the data and was involved in manuscript writing, S.B prepared the samples for TEM and performed TEM experiments, J. M A performed the analysis of the data and manuscript writing and S.D performed a complete analysis of results, supervised the whole work, planned and written the manuscript. All authors reviewed the manuscript.

## Conflicts of interest

The authors of this article declare that there are no potential conflicts of interest associated with this publication. Furthermore, there has been no significant financial support for this work that could have influenced its outcome. The authors have no relevant financial or non-financial interests to disclose. We confirm that we have carefully considered the protection of intellectual property related to this work and there are no obstacles to its publication, including any concerns regarding intellectual property timing. This manuscript does not include any materials that required ethical approval, and the research does not involve human participants and/or animals. Consequently, there is no requirement for "informed consent" in relation to this manuscript.".

## Acknowledgements

The authors acknowledge the project funding from CSIR-NAL, Bangalore.